\documentclass{article}

\usepackage{arxiv}

\usepackage[utf8]{inputenc} 
\usepackage[T1]{fontenc}    
\usepackage{hyperref}       
\usepackage{url}            
\usepackage{booktabs}       
\usepackage{amsfonts}       
\usepackage{nicefrac}       
\usepackage{microtype}      
\usepackage{lipsum}		
\usepackage{graphicx}
\usepackage{natbib}
\usepackage{doi}
\usepackage{threeparttable} 
\usepackage{amsmath,bm} 
\usepackage{bbm} 
\usepackage{subcaption} 

\title{Improving uplift model evaluation on RCT data}

\author{ \href{https://orcid.org/0000-0000-0000-0000}{\includegraphics[scale=0.06]{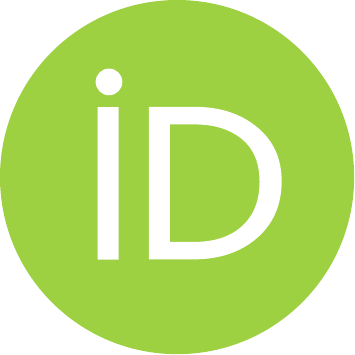}\hspace{1mm}Björn Bokelmann} \\
	Chair of information systems\\
	Humboldt University Berlin\\
	Unter den Linden 6\\
    10099 Berlin\\
	\texttt{bokelmab@hu-berlin.de} \\
	\And
	\href{https://orcid.org/0000-0000-0000-0000}{\includegraphics[scale=0.06]{orcid.pdf}\hspace{1mm}Stefan Lessmann} \\
	Chair of information systems\\
	Humboldt University Berlin\\
	Unter den Linden 6\\
    10099 Berlin\\
}

\renewcommand{\shorttitle}{Improving uplift metrics}

\hypersetup{
pdftitle={Uplift metrics},
pdfsubject={q-bio.NC, q-bio.QM},
pdfauthor={David S.~Hippocampus, Elias D.~Striatum},
pdfkeywords={First keyword, Second keyword, More},
}

\begin{document}
\maketitle

\begin{abstract}
Estimating treatment effects is one of the most challenging and important tasks of data analysts. In many applications, like online marketing and personalized medicine, treatment needs to be allocated to the individuals where it yields a high positive treatment effect. Uplift models help select the right individuals for treatment and maximize the overall treatment effect (uplift). A major  challenge in uplift modeling concerns model evaluation. Previous literature suggests methods like the Qini curve and the transformed outcome mean squared error. However, these metrics suffer from variance: their evaluations are strongly affected by random noise in the data, which renders their signals, to a certain degree, arbitrary. We theoretically analyze the variance of uplift evaluation metrics and derive possible methods of variance reduction, which are based on statistical adjustment of the outcome. We derive simple conditions under which the variance reduction methods improve the uplift evaluation metrics and empirically demonstrate their benefits on simulated and real-world data. Our paper provides strong evidence in favor of applying the suggested variance reduction procedures by default when evaluating uplift models on RCT data. 
\end{abstract}

\keywords{OR in Marketing \and Causal Machine Learning \and Uplift Modeling \and Evaluation \and Qini}


\section{Introduction}\label{introduction}
Estimating treatment effects is one of the most challenging and important tasks of data analysts. In health care, researchers infer the effect of new drugs on patients using data from clinical trials \citep{van2018proposed}. Companies estimate treatment effects from behavioral customer data to target marketing actions such as coupons or retention programs \citep{haupt2022targeting, lemmens2020managing, devriendt2021you}. In education, well-targeted interventions based on treatment effect estimates can reduce student dropout \citep{Olaya2020}. Governments also infer treatment effects to judge the adequacy of policy interventions like job training programs \citep{athey2021policy}. The application fields for treatment effect estimation are manifold and their results are crucial because healthcare organizations, companies, and governmental institutions base important decisions on such estimates. The paper focuses on estimates of conditional average treatment effects (CATE), which depend on the characteristics of individuals and facilitate personalized treatment \citep{Cousineau2023}. 

Building statistical models to support decisions on whom to treat is the objective of uplift modeling \citep{gutierrez2017causal}. There are different ways to define the modeling problem. Making treatment decisions can be seen as a "causal classification problem", where the goal is to identify individuals whose outcome positively changes by the treatment \citep{li2021general}. To make this classification problem more realistic, the cost of treatment can be included \citep{verbeke2022or}. An alternative approach is to frame the task as a ranking problem. Then, the problem becomes to rank individuals according to the effect a treatment would have on them \citep{devriendt2020learning}. An uplift model capable of ranking individuals accurately in the order of their CATE estimates helps select the individuals who profit the most from the treatment. Irrespective of how the modeling problem is defined, there are various methods for building uplift models using, for example, supervised learning algorithms \citep{devriendt2018literature} or optimization-based approaches \citep{Cousineau2023}. For this paper, the specific modeling approach is not important. It is only necessary to note that uplift models can yield estimates for the CATE 
but do not have to. It is sufficient if a model yields a score for each individual (which does not need to correspond to a CATE estimate), such that the score can be used for ranking or deciding whether to treat. Such "non-CATE models" have been shown to outperform proper CATE estimators in some applications \citep{gubela2020response, fernandez2022causal}.   
Irrespective of the uplift modeling strategy, candidate models have to be evaluated. Analysts need to appraise the precision of CATE estimates or how allocating treatment according to a model's predictions would affect business/health care. In supervised learning, analysts would use the target $Y$ from a test set and evaluate predictive accuracy using indicators like the mean squared error. Such an evaluation is impossible for uplift models. The corresponding problem is known as the fundamental problem of causal inference and concerns the fact that the modeling target, the CATE, can not be observed. What is observed are either outcomes for the treated or the untreated. Accordingly, evaluation approaches for predictive models require adjustment to support uplift modeling. The literature suggests three uplift evaluation strategies. The first, and most popular strategy, is to evaluate how well an uplift model ranks individuals according to their CATE. This works by cutting the test set observations into segments according to the model predictions (e.g. the individuals with the 10\% highest predictions) and then calculating the mean difference between the treated and the untreated in this segment. If the model ranks well, the observed differences in a high-ranked segment will also be high. This approach, and some variations of it, is found under the name \textit{Qini} or \textit{uplift curve} in the literature \citep{devriendt2020learning}. Qini or uplift curves have a clear economic interpretation. They show for each campaign size (number of treated individuals) the economic gain if treatment is allocated according to the uplift model. A second evaluation strategy is to measure how accurate an uplift model's predictions are for the actual CATE. The literature suggests different metrics, measuring the squared difference between CATE estimates and some substitutes of the actual (unobservable) CATE \citep{gutierrez2017causal, schuler2018comparison, saito2020counterfactual}. Although such accuracy metrics are far less popular in the uplift literature than ranking metrics, we argue that they are an important complement. If the treatment costs differ between individuals, pure ranking ability of a model is not sufficient to support targeted treatment decisions \citep{haupt2022targeting}. In such application settings, accuracy of CATE predictions becomes important. The third evaluation strategy goes one step further than the ranking and accuracy metrics by evaluating not the predictions of the model, but already the decisions derived from it \citep{zhao2017uplift}. These \textit{decision metrics} measure, for example, how the mean observed outcome changes if one would allocate treatment according to model recommendations. Our research focuses on the ranking and accuracy metrics, but we also show that the obtained results can directly be transferred to decision metrics.        

Uplift model evaluation is the topic of this paper. To discuss properties of uplift evaluation metrics, we apply the well known concept of \textit{bias and variance} from predictive modeling in the context of evaluation metrics. In this context, a metric would be biased, if it favors any kind of uplift model over another, unrelated to their real performance (which we will discus in more detail later). The magnitude of variance describes the extend to which model evaluations depend on the randomness in the sampling of the test data. A high variance of an evaluation metric means that a large test set size is required to obtain meaningful model evaluations. 

To see where our research improves current uplift model evaluation routines, we need to consider two peculiarities of the uplift literature. First, uplift modeling is often done using data from \textit{randomized controlled trials} (RCT) \citep{ICIS2019}. Random  treatment allocation eliminates potential confounders and thereby renders the Qini or uplift curve unbiased. However, our research shows that traditional Qini or uplift curves, which are calculated by differences in means of outcomes, are sub-optimal regarding the variance. We show that calculating these metrics based on suitable \textit{adjusted outcomes} leads to a variance reduction. This result implies that many uplift studies have not optimally used data. More specifically, the same certainty in model evaluations could have been achieved with less (costly) test data, \textit{if} a study had used our proposed outcome adjustment methods. For example, recent studies that could directly benefit from our propositions include \citet{verbeke2022or, berrevoets2022treatment, de2021uplift, devriendt2021you, baier2022profit, betlei2021uplift}; amongst many others. We highlight that, in practice, the variance reduction by outcome adjustment is relevant. Gathering RCT data is costly and accordingly a reduction of the required sample size yields economic benefits. 
A second characteristic of uplift evaluation is that accuracy metrics are rarely applied. As we will show, the few  exceptions like \citep{gutierrez2017causal} and \citep{hitsch2018heterogeneous}, which exemplify the use of accuracy measures for uplift modeling, propose measures that are unbiased but sub-optimal regarding variance. Again, our outcome adjustment improves these measures.      

In summary, the paper contributes to the uplift literature by theoretically proving that suitable outcome adjustments for uplift metrics on RCT data can reduce their variance, while keeping them unbiased. We also derive clear conditions under which a variance reduction is achieved. Based on these conditions, we provide three different outcome adjustment methods for reducing the variance. We apply our outcome adjustment methods in a simulation study and on three real-world data sets, and consistently observe the outcome adjustment methods to reduce the variance of an evaluation metric without introducing bias. The theoretical and empirical analysis provide clear evidence that uplift evaluation procedures on RCT data can be improved by applying the suggested adjustment methods. The empirical results further suggest that outcome adjustment can reduce the required size of the testing data by 10\%, which we consider managerially meaningful. The complete code of our empirical analysis is provided by \citet{bokelmab}. 

\section{Related literature}
In this section, we review related literature on uplift model evaluation. In doing so, we deliberately leave out the vast literature dealing with uplift model estimation. A summary of the many available methods is available in, for example, \citet{gutierrez2017causal, jacob2021cate, Cousineau2023}. Our analysis starts after model estimation. We focus on the question of how to best evaluate the given model(s) on a test set. Our paper aims to improve uplift model evaluation on RCT data. While our research is a contribution to the \textit{uplift literature}, it is methodically related to another literature branch, namely \textit{causal inference}. By mainly exploring how to extract treatment effects from observational data, where treatment allocation is not random, causal inference has a different focus than uplift modeling. However, there are causal inference studies dealing with the CATE estimation, and as the CATE is also the target of uplift modeling, there is a relationship between both literature streams \citep{gutierrez2017causal}. Therefore, to comprehensively place our research findings in the context of the previous literature, we describe related concepts and studies from the causal inference literature in detail as well. We start our review by describing two concepts originating from the causal inference literature, which are important for our research and, to our best knowledge, not yet considered in state-of-the-art uplift evaluation routines. Thereafter, we describe the uplift and causal inference literature relating to ranking and accuracy metrics and show where our research extends and complements the corresponding research. 

The first important concept from the causal inference literature is the \textit{doubly-robust estimation procedure}, an effect estimation procedure for observational data, tending to generate estimates with low bias and variance \citep{robins1995semiparametric,chernozhukov2018double}. The relationship to our research is that one of our proposed outcome adjustment methods turns out to be similar to the doubly-robust estimation procedure on RCT data. Related to our research, the doubly-robust procedure has already been used to estimate CATE models \citep{chernozhukov2018double}, but also to adapt CATE model metrics to the case of observational data \citep{saito2020doubly}. In the causal inference literature, it is also known that the doubly-robust estimation procedure can reduce the variance of effect estimates on RCT data, but brings the potential problem of bias because the estimation procedure involves building statistical models (which are generally biased). Prevention of biased effect estimates is the purpose of \textit{cross-fitting}. The theoretical result of cross-fitting guarantees that certain model-based effect estimates, like the doubly-robust estimates, are unbiased on RCT data, if models are build on separate data than they are used to estimate effects \citep{wager2016high}. Apart from CATE model estimation \citep{chernozhukov2018double}, these two concepts have already been combined to reduce the variance of ATE estimates on RCT data \citep{wager2016high,guo2021machine,jin2022towards}. 

In the following, we describe the current state-of-art in uplift model evaluation and how our research leads to new insights and improvements. In terms of ranking metrics, our paper builds on \citet{radcliffe2007using}, who suggests the Qini curve for uplift model evaluation. Since then, the Qini curve, and slight modifications of it, are the most widely applied evaluation metrics for uplift models. 
Recent work by \citet{devriendt2020learning} provides a comprehensive overview of different Qini curve versions applied thus far. Albeit some variations, the basic idea remains the same: treatment effects are calculated by subtracting (unadjusted) outcomes of the treated from the untreated. By suggesting to \textit{adjust the outcome}, our paper extends previous works on ranking metrics for uplift evaluation. While completely new in the uplift literature, ranking metric calculation procedures by adjusted outcomes have already been discussed in a few causal inference studies. To our knowledge, the first authors using some outcome adjustment (specifically doubly-robust estimation) to calculate the Qini-curve are \citet{saito2019doubly}. They applied this estimation procedure to use the Qini curve on observational data, where the version of \citet{radcliffe2007using} would suffer from confounding. However, their paper does not discuss the possible use of doubly-robust estimation for ranking metrics on \textit{RCT data}. In fact, in contrast to the traditional way of calculating the Qini curve \citep{radcliffe2007using}, such estimates would be biased. To the best of our knowledge, only \citet{yadlowsky2021evaluating} have noted the potential of doubly-robust estimation for the Qini curve on RCT data and found a way to prevent the bias by cross-fitting.       

To see where our research complements previous causal inference literature on ranking metrics, it is necessary to describe the study by \citet{yadlowsky2021evaluating} in more detail. In this very general study, the authors derive asymptotic distribution properties and inference methods for multiple ranking metrics in the context of CATE estimation. Thereby, they analyze the case of observational data, survival data but also RCT data. Within this scope, they state that, on RCT data, doubly-robust estimation "can help to reduce" the variance of the Qini curve for uplift modeling, compared to the traditional calculation procedure by \citet{radcliffe2007using}. However, they do not analyze under which conditions the doubly-robust estimation leads to variance reduction for finite test sets. They also do not analyze whether other statistical procedures lead to (possibly even more) variance reduction and they do not empirically examine by how much doubly-robust (or other methods) reduce the variance compared to the usual procedure by \citet{radcliffe2007using}, therefore leaving unanswered the question of practical relevance. Regarding the Qini curve, our paper contributes to the literature by answering all these questions.

In terms of accuracy metrics, our research builds on the work of \citet{gutierrez2017causal} and \citet{hitsch2018heterogeneous}, who propose the \textit{transformed outcome mean squared error}, which we denote by $MSE_{W}$, as an accuracy metric for uplift. Subsequent papers stem from the causal inference literature and address the evaluation of CATE model's accuracy on observational data. \citet{schuler2018comparison}, \citet{saito2019doubly} and \citet{saito2020counterfactual} added the plug-in loss $MSE_{pi}$, the $\tau$-risk $MSE_{\tau}$ and the $\mu$-loss $MSE_{\mu}$ as alternative measures to the $MSE_{W}$. These measures were inspired by loss functions previously used for estimating CATE models. \citet{alaa2019validating} suggest a way to update the $MSE_{\tau}$ by influence functions to reduce the bias. The most recent study comes from \citet{mahajan2022empirical}, who empirically assess all the above mentioned accuracy metrics. The results do not show superiority of any of the suggested metrics, although the $MSE_{\tau}$ performed best on average. In summary, \textit{on observational data}, there is a great variety of previously suggested accuracy metrics. Both  empirical \citep{schuler2018comparison} and theoretical \citep{saito2020counterfactual} results indicate that the classical $MSE_{W}$ is inferior to other accuracy metrics. However, there is no clear guidance on which metric to use. 

Regarding accuracy metrics, our paper complements the above studies by thoroughly examining the case of \textit{RCT data}. This allows us to provide clear guidelines which metric to apply. Although previous causal inference studies already theoretically analyze bias and variance of accuracy metrics \citep{saito2020counterfactual}, we obtain results for the RCT case that are not yet derived or discussed: First, unlike the $MSE_{W}$, most of the other suggested accuracy metrics are biased on RCT data. This limits the range of suitable metrics to the $MSE_{W}$ and some \textit{outcome-adjusted} versions of it. Second, we derive conditions under which these \textit{outcome-adjusted} versions of the $MSE_{W}$ have lower variance than the classical $MSE_{W}$. Based on these theoretical results and an empirical study, we provide clear advice how to evaluate uplift model accuracy on RCT data.

\section{Evaluation of uplift models}
In this paper, we consider the following situation: An RCT was performed and data was split into a training and a test set. Uplift models $\hat{\tau}^{(1)}_{x},\hat{\tau}^{(2)}_{x},...$ were built on the training set and are to be evaluated on the test set. We consider a single treatment and aim at measuring its effect over a no-treatment (i.e., control) case. The treatment was assigned randomly with fixed probability $p$. The outcome is of the form
\begin{align}
    Y=\mu_{x}+W\cdot \tau_{x}+\varepsilon\label{eq_dec_Y},
\end{align}
where $W \in \left \{ 0,1 \right \} $ indicates whether treatment was assigned, $\mu_{x}$ is the conditional expectation of $Y$ given $X$ in the control group, $\tau_{x}$ is the CATE, and $\varepsilon$ is the noise which can not be explained by $X$. 

The question to be answered is "how can we measure the performance of the models $\hat{\tau}^{(1)}_{x}$, $\hat{\tau}^{(2)}_{x}$, $...$?". Previous literature suggests either evaluating the accuracy of model predictions $\hat{\tau}_{x}$ for the true $\tau_{x}$ or evaluating how well a model can rank individuals according to $\tau_{x}$ \citep{gutierrez2017causal}. The literature also suggests how to calculate this performance on the test set. We call the corresponding performance calculations on a test set "empirical evaluation metrics". Before we introduce the suggested empirical evaluation metrics, we first analyze uplift model performance theoretically. This is important because we will show that the empirical evaluation metrics always mean to estimate a theoretical performance measure based on the test set. 

With the distinction between theoretical performance measures and empirical evaluation metrics, it becomes clear what the familiar concepts of bias and variance mean \textit{in the context of evaluation metrics}. A bias is a systematical deviation of the empirical evaluation metric from the theoretical performance measure. If an empirical evaluation metric is unbiased, it converges to the theoretical performance measure with a growing test set size. For test sets of finite sample size, there is some random variation of the empirical evaluation metric around the theoretical performance measure. This random variation is the variance of the empirical evaluation metric. 

In the next subsection, we provide the theoretical performance measures, which the empirical evaluation metrics mean to estimate. After that, the next two subsections introduce the empirical evaluation metrics provided in prior literature. Subsection \ref{sec_tomse} provides the empirical evaluation metric to assess the accuracy of model estimates $\hat{\tau}_{x}$ for the true $\tau_{x}$. Subsection \ref{sec_rank_metr} provides the empirical evaluation metric to assess how well a model can rank individuals according to $\tau_{x}$.    

\subsection{Theoretical uplift model performance}\label{sec_def_perf}
We start with discussing accuracy: Here, previous literature already defines theoretical performance by the mean squared error 
\begin{align}
MSE(\hat{\tau}_{x})=E[(\tau_{x}-\hat{\tau}_{x})^{2}]\label{eq_mse}
\end{align}\citep{gutierrez2017causal, hitsch2018heterogeneous}. Note that this performance measure only applies to uplift models that yield CATE estimates. 

\begin{figure}[!htb]
\centering
  \includegraphics[width=0.6\textwidth]{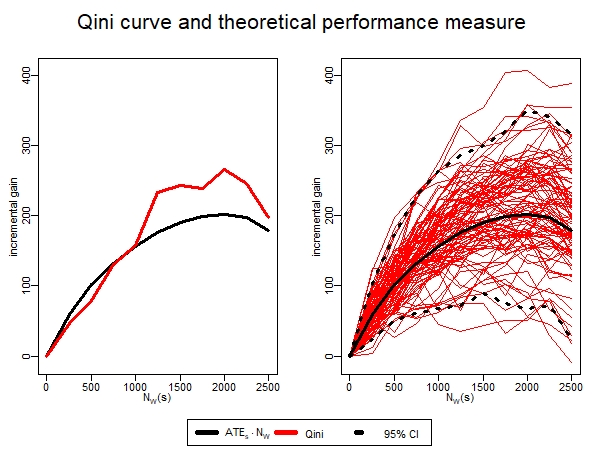}
  \captionof{figure}{Theoretical ranking performance against the empirically calculated Qini curve. The left plot shows the theoretical performance $ATE_{s}\cdot N_{W}$ in comparison to the Qini curve calculated on a single simulated test set according to setting "nw" in section \ref{data}. The right plot shows the same, only for 100 simulated test sets. The dashed lines represent the highest/lowest 2.5-percentile of Qini curves at each point.}
  \label{fig:Qini_plot}
\end{figure}

For ranking performance, previous literature mostly focuses on how to calculate the empirical metric on the test set, instead of analysing what this measures theoretically. Most suggested empirical evaluation metrics go back to \citet{radcliffe2007using}, whose idea of ranking performance can be described by the following thought experiment: The model is provided a data set with individuals and their respective features. It is allowed to select a share $s$ of these individuals that receive treatment. The model should select  individuals such that the incremental gain (overall treatment effect) is maximized. A model which is good at ranking according to $\tau_{x}$ will be able to select the right individuals and achieve a high incremental gain. Based on this notion of performance, \citet{radcliffe2007using} suggests the Qini curve, which shows estimates of the incremental gain for each share $s$. Theoretically, we can express this incremental gain for each share $s$ by   
\begin{align}
g_{s}(\hat{\tau}_{x})&=ATE_{s}(\hat{\tau}_{x})\cdot N_{W}\label{eq_ates},
\end{align} where $ATE_{s}(\hat{\tau}_{x}):=E[\tau_{x}]$ is the average treatment effect of the share $s$ highest ranked individuals and $N_{W}$ is the number of treated individuals within this share. In section \ref{sec_rank_metr} and \ref{ap_other_metrics}, we describe the relationship between this theoretical performance measure and the calculation rule by \citet{radcliffe2007using} in more detail. Figure \ref{fig:Qini_plot} depicts a Qini curve calculated on a test set in relation to the theoretical performance measure.        
While the theoretical performance measures are intuitive, a practical challenge is that we do not observe $\tau_{x}$ in the test set. Hence, we cannot calculate $MSE(\hat{\tau}_{x})$ and $ATE_{s}(\hat{\tau}_{x})$. Instead, the literature suggests ways to estimate $\tau_{x}$ and use these estimates in the above definitions of the performance measures \eqref{eq_mse} and \eqref{eq_ates}. It is this estimation of $\tau_{x}$ where variance enters the evaluation metrics. Instead of getting performance values \eqref{eq_mse} or \eqref{eq_ates}, the obtained evaluation metrics are only estimates for them. These estimates vary across observations in the test set and thus across test sets. 

In the next two sections, we show how the evaluation metrics are calculated in practice. Thereby, we derive a simple mathematical expression for the variance in the evaluation metrics. Given this expression, it will be easy to see that current uplift model evaluation practices suffer from unnecessarily high variance and how the variance of the metrics can be reduced.  

\subsection{Transformed outcome mean squared error}\label{sec_tomse}
The transformed outcome mean squared error is an approach to evaluate uplift models according to the mean squared error loss defined in equation \eqref{eq_mse}. \citet{gutierrez2017causal} and \cite{hitsch2018heterogeneous} suggest to replace $\tau_{x}$ by the transformed outcome $W^{p}Y$, where $W^{p}$ is the \textit{Horvitz-Thompson transformation}, defined by  
\begin{align}
    W^{p}:=\frac{W}{p}-\frac{1-W}{1-p}\label{eq_thtrans}.
\end{align} We denote the resulting \textit{transformed outcome mean squared error} by
\begin{align}
MSE_{W}(\hat{\tau}):=\frac{1}{N}\sum (W^{p}Y_{i}-\hat{\tau}_{x_{i}})^{2}.\label{eq_MSEto}
\end{align} In previous literature, this metric is applied to evaluate model predictions on RCT data \citep{hitsch2018heterogeneous,haupt2022targeting}. For observational data, the same metric could be applied with the only difference that the constant $p$ in $W^{p}$ needs to be replaced by a propensity score $\hat{p}(x_{i})$, estimating the treatment probability of each individual $i$. 

The causal inference literature suggests some alternative empirical evaluation metrics to estimate the theoretical accuracy performance \eqref{eq_mse} \citep{saito2020counterfactual,schuler2018comparison}. However, in \ref{app_alt_acc}, we show that except for the $MSE_{W}$ (and outcome adjusted versions of it), these metrics are biased on RCT data. Therefore, we restrict the following analysis to the $MSE_{W}$.

Now, we analyze the expected value and variance of $MSE_{W}(\hat{\tau})$. For the expected value, it is useful to analyze the difference in $MSE_{W}$ between two uplift models because the expected value of the $MSE_{W}$ for a single model entails some hard-to-interpret expressions, which would make our discussion unnecessarily complex \citep{gutierrez2017causal}. In \ref{app_distr_TOMSE}, we derive that the expected value of the difference in $MSE_{W}$ between two models is given by
\begin{align}
E[MSE^{\hat{\Phi}}_{W}(\hat{\tau}^{(1)}_{x})-MSE^{\hat{\Phi}}_{W}(\hat{\tau}^{(2)}_{x})]=E[(\tau_{x}-\hat{\tau}^{(1)}_{x})^{2}]-E[(\tau_{x}-\hat{\tau}^{(2)}_{x})^{2}]\label{eq_mse_unb}.
\end{align} This result is reassuring as it shows that the evaluation is unbiased. In expectation, the difference in $MSE_{W}$ is the difference in the theoretical MSE from equation \eqref{eq_mse}. 

For infinite sample size, the empirical evaluation metric would exactly correspond to the theoretical performance in equation \eqref{eq_mse_unb}. However, as the (test) sample size is always finite in practice, we derive the variance of the difference in $MSE_{W}$ in \ref{app_distr_TOMSE}. Since this analysis is quite complex, we rather present only the essence of our analysis here. The important result is that the variance of the difference in $MSE_{W}$ depends on the variance of the transformed outcome $W^{p}Y$ and certain irreducible components. Using the law of total variance, the variance of the transformed outcome can be decomposed to
\begin{align}
Var[W^{p}Y]=E[Var[W^{p}Y|W]]+Var[E[W^{p}Y|W]]\label{eq_dec_wpY}.
\end{align} Any methods striving for a variance reduction of the $MSE_{W}$ thus need to reduce the two components on the right side of the above equation. 

We close this section by providing some intuition about this result. The idea behind the $MSE_{W}$ is to replace the true CATE $\tau_{x}$ by an unbiased estimator - the transformed outcome $W^{p}Y$. This transformed outcome varies around $\tau_{x}$ and this variance translates into the variance of the $MSE_{W}$. So, it is intuitive that any potentially reducible variance in the $MSE_{W}$ is due to the variance of $W^{p}Y$. 

\subsection{Qini curve}\label{sec_rank_metr}
The Qini curve is an approach to empirically evaluate uplift models according to their ranking ability on the test set. Many slight variations of the Qini curve have appeared in the literature \citep{devriendt2020learning}. There are also other names like "uplift curve", which basically refer to the same metric. In the following, we limit our analysis to the Qini curve suggested by \citet{radcliffe2007using}. Importantly, the results of our analysis can be transferred to other versions of the Qini curve, \ref{ap_other_metrics} discusses these other forms in more detail.

For his Qini curve, \citet{radcliffe2007using} measures, for each share $s$ of highest ranked individuals, by how much the outcome increased due to the treatment. He calls this the "incremental gain", which he calculates by
\begin{align*}
u_{s}=\sum Y_{i}|_{W=1}-\frac{N_{W}}{N_{\bar{W}}}\sum Y_{i}|_{W=0},
\end{align*} where $N_{W}$ is the number of treated individuals within the share $s$ of the highest ranked individuals in the test set and $N_{\bar{W}}$ is the number of untreated individuals within the share $s$ of the highest ranked individuals in the test set. $Y_{i}|_{W=1}$ and $Y_{i}|_{W=0}$ refer to the outcomes of the treated respectively untreated individuals. The Qini curve is then just a plot of $u_{s}$ for selected shares $s$. 

To see how this Qini curve relates to the theoretical performance measure of equation \eqref{eq_ates}, we write down an empirical estimator of the average treatment effect for the share $s$ highest ranked individuals  
\begin{align}
\hat{ATE}_{s}(\hat{\tau}_{x})=\frac{1}{N_{W}}\sum Y_{i}|_{W=1}-\frac{1}{N_{\bar{W}}}\sum Y_{i}|_{W=0}\label{eq_qini}.
\end{align} We note that the qini curve $u_{s}$ can then be written as
\begin{align*}
u_{s}=\hat{ATE}_{s}(\hat{\tau}_{x})\cdot N_{W}.
\end{align*} So, the Qini curve calculated on the test set can be seen as an estimate for the theoretical ranking performance $ATE_{s}\cdot N_{W}$. The relationship between $u_{s}$ and the theoretical performance measure is illustrated in Figure \ref{fig:Qini_plot}. 

Next, we discuss bias and variance of the Qini curve. For these statistical properties, the factor $N_{W}$ in the Qini curve is irrelevant. It is sufficient to examine the bias and variance of $\hat{ATE}_{s}(\hat{\tau}_{x})$ as an estimator for $ATE_{s}(\hat{\tau}_{x})$. From the definition of $\hat{ATE}_{s}(\hat{\tau}_{x})$ as the difference in sample means, it is clear that $\hat{ATE}_{s}(\hat{\tau}_{x})$ is unbiased (see \ref{ap_qini_distr}), that is
\begin{align}
    E[\hat{ATE}_{s}(\hat{\tau}_{x})]=E[\tau_{x}],\label{eq_qini_unb}
\end{align} where $\tau_{x}$ belongs to the individuals in the highest ranked share $s$. For the variance, we derive in \ref{ap_qini_distr} 
\begin{align*}
Var[\hat{ATE}_{s}(\hat{\tau}_{x})]=\frac{1}{N}E[Var[W^{\tilde{p}}Y|W]],
\end{align*} where the outcome $Y$ on the right side of the equation belongs to the share $s$ of the highest-ranked individuals. What is important here is that the metrics variance depends on $E[Var[W^{\tilde{p}}Y|W]]$. This is one of the two variance components of $Var[W^{p}Y]$ in equation \eqref{eq_dec_wpY}, only with $\tilde{p}$ being the fraction of treated within the share $s$, instead of the treatment probability $p$ in the RCT.      

\section{Metric variance reduction by outcome adjustment}\label{sec_met_var_red}
\subsection{The principle of variance reduction}
In our analysis of the uplift evaluation metrics, we found that their variance depends on the variance of $W^{p}Y$. Accordingly, it is worthwhile to analyze the variance of $W^{p}Y$ in detail and examine whether it could be reduced. For this purpose, it is useful to write $W^{p}Y$ in the following form
\begin{align}
W^{p}Y=\tau_{x}+W^{p}\Phi(X)+W^{p}\varepsilon\label{eq_WpY},
\end{align}
where $\Phi(X):=\mu_{x}+(1-p)\tau_{x}$ is called the \textit{nuisance function} in the following. Given this equation, we can easily explain the idea of variance reduction: The purpose of $W^{p}Y$ in the evaluation metrics is always to replace the unobservable $\tau_{x}$. Everything except $\tau_{x}$ on the right side of equation \eqref{eq_WpY} only causes variance of $W^{p}Y$ around $\tau_{x}$, and this variance will enter the evaluation metrics. Hence, it would be ideal to remove $W^{p}\Phi(X)$ and $W^{p}\varepsilon$. However, $W^{p}\varepsilon$ is irreducible, because it is due to the unexplainable noise. But if we could remove $W^{p}\Phi(X)$, this would already lead to a reduction of variance. The idea behind outcome adjustment is simply to train a supervised learning model $\hat{\Phi}(X)$ for the nuisance function $\Phi(X)$ on the training data and adjust the outcome to $Y-\hat{\Phi}(X)$ on the test data. Then, the reducible noise component becomes $W^{p}(\Phi(X)-\hat{\Phi}(X))$. In the following, we examine under which conditions such an outcome adjustment leads to a variance reduction of $W^{p}Y$ and thereby of the evaluation metrics.

First, we want to examine in more detail how the nuisance function $\Phi(X)$ affects the variance of the uplift evaluation metrics. Remember that the variance of $W^{p}Y$ can be decomposed into two components (see equation \eqref{eq_dec_wpY}). In \ref{app_calc_var_comp}, we derive the following equations for these components\begin{align}
E[Var[W^{p}Y|W]]&=Var[\tau_{x}]+\frac{Var[\Phi(X)]+Var[\varepsilon]}{p(1-p)}\label{eq_var_comp_e} \text{ and}\\
Var[E[W^{p}Y|W]]&=\frac{E[\Phi(X)]^{2}}{p(1-p)}\label{eq_var_comp_v}.
\end{align} These equations clearly show how the nuisance function affects the variance components of $W^{p}Y$. In section \ref{sec_tomse}, we have seen that both components determine the variance of the $MSE_{W}$. In section \ref{sec_rank_metr}, we have seen that only the first component affects the Qini curve. This leads to the following conditions for variance reduction of the evaluation metrics: If an adjustment function $\hat{\Phi}(X)$ fulfills
\begin{align*}
|E[\Phi(X)]-E[\hat{\Phi}(X)]|<|E[\Phi(X)]|\label{eq_cond1} \tag{C1},
\end{align*} this leads to a variance reduction of the $MSE_{W}$. If the adjustment function further fulfills    
\begin{align*}
Var[\Phi(X)-\hat{\Phi}(X)]<Var[\Phi(X)]\label{eq_cond2} \tag{C2},
\end{align*} this leads to further variance reduction of the $MSE_{W}$ and additionally reduces the variance of the Qini curve. The conditions \eqref{eq_cond1} and \eqref{eq_cond2} lead directly to plausible choices of adjustment functions $\hat{\Phi}(X)$. 

\subsection{Outcome adjustment methods}\label{sec_adj_methods}
In this section, we  provide adjustment methods that fulfill conditions \eqref{eq_cond1} and \eqref{eq_cond2} and thereby reduce the variance of the evaluation metrics. We start with condition \eqref{eq_cond1}. This condition is already fulfilled if we construct a suitable estimate of $E[\Phi(X)]$. Using the outcome representation \eqref{eq_dec_Y}, we can verify that
\begin{align*}
E[\Phi(X)]=(1-p)E[Y|W=1]+pE[Y|W=0].
\end{align*} So, an easy choice for an adjustment function would be the \textit{unconditional mean adjustment}
\begin{align*}
\hat{\Phi}^{uc}(X)&:=(1-p)\hat{\mu}_{1}+p\hat{\mu}_{0}\tag{AM1}\label{eq_am1},
\end{align*} where $\hat{\mu}_{1}, \hat{\mu}_{0}$ are respectively the sample averages of the treated and untreated in the training set. Note that the unconditional mean adjustment is constant in $X$. This is why it has no effect on condition \eqref{eq_cond2}. So, using this adjustment method, the variance of the Qini curve remains unaffected. 

Next, we examine how an adjustment method could fulfill condition \eqref{eq_cond2}. Again, we can use the outcome representation \eqref{eq_dec_Y} to verify that
\begin{align*}
\Phi(X)=(1-p)E[Y|W=1,X]+pE[Y|W=0,X].
\end{align*} The conditional expected values on the right-hand side are the targets of supervised learning models for the outcome of the treated respectively untreated individuals. So, a natural choice for the adjustment function would be 
\begin{align*}
\hat{\Phi}^{DR}(X)&:=(1-p)\cdot\hat{\mu}_{1}(x)+p\cdot \hat{\mu}_{0}(x)\tag{AM2}\label{eq_am2},
\end{align*} where $\hat{\mu}_{1}(x), \hat{\mu}_{0}(x)$ are supervised learning models trained on the treated respectively untreated individuals in the training set. We call this method \textit{doubly-robust adjustment} because we can show that $W^{p}(Y-\hat{\Phi}^{DR}(X))$ corresponds to the famous doubly robust outcome transformation \citep{robins1995semiparametric} adapted to RCT data (that means $p$ is a fixed value instead of a propensity score). It can be shown that the doubly-robust method \citet{yadlowsky2021evaluating} apply to estimate the Qini curve on RCT data is equivalent to this doubly-robust outcome adjustment combined with the traditional way of \citet{radcliffe2007using} to calculate the Qini curve. Furthermore, the doubly-robust version of the $MSE_{W}$, which \citet{saito2020counterfactual} analyze on observational data is similar to $\hat{\Phi}^{DR}$, only with an estimated propensity score $\hat{p}(x)$ instead of a fixed probability $p$. In the most common RCT case, where $p=0.5$, another adjustment method is plausible. This is because in this case:
\begin{align*}
\Phi(X)=E[Y|X].
\end{align*}
This makes \textit{conditional outcome adjustment} a plausible method. The adjustment function is given by 
\begin{align}
\hat{\Phi}^{c}(X)&:=\hat{\mu}(x)\tag{AM3}\label{eq_am3},
\end{align} with $\hat{\mu}(x)$ denoting a supervised learning model trained on the whole training data (treated and untreated individuals combined).    

\subsection{Practical impact of the adjustment methods}
In the previous sections, we derived adjustment methods from theoretical results regarding the variance of uplift evaluation methods. Next, we analyze what is expected to happen if we apply them in practice. Specifically, we will deal with three questions: \textit{Can outcome adjustment bias uplift evaluation metrics? Does outcome adjustment always reduce the variance of an uplift evaluation metric or could it even increase variance? What is the practical impact of variance reduction?} 

We start with bias. Interestingly, the adjustment methods can not introduce bias in the evaluation metrics simply because the adjustment functions $\hat{\Phi}(X)$ are fitted on the training set and not on the test set, where they are used for adjustment. On the test set, $\hat{\Phi}(X)$ is just a fixed function. In consequence, the outcome in equation \eqref{eq_dec_Y} becomes 
\begin{align*}
Y-\hat{\Phi}(X)=\bar{\mu}_{x}+W\cdot\tau_{x}+\varepsilon
\end{align*} with $\bar{\mu}_{x}=\mu_{x}-\hat{\Phi}(X)$. In the equations \eqref{eq_mse_unb} and \eqref{eq_qini_unb} for the expected value of the $MSE_{W}$ and the Qini curve, the conditional expected value $\mu_{x}$ of the outcome does not appear. Accordingly, these expected values are unaffected by the shift in conditional expected value to $\bar{\mu}_{x}$. Hence, the metrics remain unbiased by outcome adjustment. We highlight that this only holds if the adjustment functions are not trained on the test set. The same principle of unbiasedness by doing model training and effect estimation on different data sets applies in previous literature about CATE model training under the name "cross fitting" \citep{chernozhukov2018double} or "honest trees" \citep{wager2018estimation}.

Next, we discuss whether outcome adjustment always leads to variance reduction. This equates to asking whether conditions \eqref{eq_cond1} and \eqref{eq_cond2} always hold for the suggested adjustment methods. In \ref{app_red_vc}, we provide a theoretical argument why it is almost impossible for the adjustment methods to fail condition \eqref{eq_cond1} unless $E[\Phi(X)]$ is almost zero. Regarding condition \eqref{eq_cond2}, we already noted that unconditional mean adjustment has no effect. For the doubly-robust and the conditional mean adjustment, we show that condition \eqref{eq_cond2} is fulfilled if their adjustment functions approximate $\Phi(X)$ better than a naive model ($\hat{\Phi}(X)=E[\Phi(X)]$). So, if the underlying supervised learning models of the adjustment methods are not fitted very poorly, condition \eqref{eq_cond2} holds. In summary, if the training of the adjustment functions is done properly, there will, except for very special settings ($E[\Phi(X)]\approx 0$ and $Var[\Phi(X)]\approx 0$), be a reduction of the metrics variance. However, the magnitude of this effect is not yet clear. 

Last, we discuss the impact of variance reduction. In practice, analysts have a test set on which they evaluate uplift model performance. On this test set, the variance of an evaluation metric would be reflected in the width of the confidence interval of an evaluation metric. Clearly, wide confidence intervals indicate that the model evaluation offers little certainty about the real (theoretical) performance of the model. In \ref{app_CI}, we derive formulas for confidence intervals of the $MSE_{W}$ and the Qini curve. From these formulas, an important fact can be derived: The width of the confidence interval is proportional to the square root of the variance of the error metrics and anti-proportional to the sample size of test set. This implies that reducing the variance of a metric by outcome adjustment allows for a reduction of the test set sample size without affecting the certainty of evaluation (width of confidence interval). Therefore, a reduction of variance by x\% can be understood as a reduction of the required test set size by x\%. In the following empirical analysis, we will measure the magnitude of this reduction.

\section{Empirical analysis}
In this section, we empirically assess the impact of the outcome adjustment methods. We provide the results of an experiment with simulated data as well as an experiment with three real-world data sets. The complete code for the empirical analysis is publicly available at \citep{bokelmab}. 

\subsection{Data}
\subsubsection{Simulated data}\label{data}
For two simulation settings, we made 10.000 simulation runs. In each of the 10.000 simulation runs, we generated 15.000 samples as a data set and used 10.000 observations for training and 5.000 observations for evaluation. We choose simulated data from two previously published studies using the R function "generate\_causal\_data" from the grf package \citep{grf_pack}. Each of the settings simulates the features $X_{i}$ and the treatment group $W\sim Bern(0.5)$, as well as the outcome of the form
\begin{align*}
Y&=\bar{\mu}_{x}+(W-0.5)\bar{\tau}_{x}+\varepsilon \\
\bar{\mu}_{x}&=a_{x}/\sigma(a_{x})\\
\bar{\tau}_{x}&=b_{x}/\sigma(b_{x})*0.1,
\end{align*} where $\varepsilon\sim N(0,\sigma^{2})$. We always used 6 features $X$. We choose three different values of the error standard deviation $\sigma \in\{0.5, 1, 2\}$. 

The two simulation settings are taken from, respectively, \citet{wager2018estimation} and \citet{nie2021quasi}, and are denoted by "aw" and "nw" in the following. The simulation parameters for aw are
\begin{align*}
a_{x}&=0.5\left(1+\frac{1}{1+e^{-20(X_{1}-1/3)}}\right)\left(1+\frac{1}{1+e^{-20(X_{2}-1/3)}}\right)\\
b_{x}&=\left(1+\frac{1}{1+e^{-20(X_{1}-1/3)}}\right)\left(1+\frac{1}{1+e^{-20(X_{2}-1/3)}}\right)\\
X_{i}&\sim U(0,1).
\end{align*}

The "nw" simulation parameters are
\begin{align*}
a_{x}&=max(0,X_{1}+X_{2},X_{3})+max(0,X_{4}+X_{5})+0.5(X_{1}+log(1+exp(X_{2}))\\
b_{x}&=X_{1}+log(1+exp(X_{2})\\
X_{i}&\sim N(0,1).
\end{align*}

\subsection{Real-world data}
For our real-world data analysis, we use three data sets: The publicly available Criteo data set provided by \citet{Diemert2018}, the publicly available Hillstrom data set provided by \citet{hillstrom}, and a data set related to digital couponing, which was provided by an online retailer \citep{haupt2022targeting, gubela2020response}. We refer to the latter data set as "the private data set". 

All of these data sets contain data of RCT's. The Criteo data set contains 13,979,592 observations and one randomized treatment with a treatment probability of $p=0.85$. The Hillstrom data set contains two different treatments (advertisement for women's or men's clothes), of which we deleted the observations with an advertisement for men's clothes, to obtain an RCT with one treatment and a treatment probability of $p=0.5$ and 42,693 observations. The private data set contains 144,630 observations with one treatment and a treatment probability of $p=0.75$.

From the Hillstrom and the private data set, we used 80\% of the data for training and 20\% of the data for testing. The Criteo data set, with its very large sample size, posed some computational challenges. From this data set, we used 90\% of the observations for testing and the remaining observations for training. 

\subsection{Models for uplift and outcome adjustment}
The actual choice of an uplift model to evaluate is of little importance for the empirical analysis. We do not want to find out which uplift modeling method is the best, but rather how to reliably evaluate an uplift model (no matter which). The only point to consider when choosing an uplift model was that the $MSE_{W}$ is only applicable to a model that yields CATE predictions. We choose a causal random forest $\hat{\tau}_{x}^{CF}$ for our analysis \citep{athey2019generalized}. Tuning parameters were fixed according to run time aspects: For the computationally intensive simulation, we fixed 1.000 trees and took the default parameters from the \textit{causal\_forest}-function \citep{grf_pack}. For the large Criteo data set, we also fixed 1.000 trees and further reduced run time by fitting the model on a subset of size 100,000 from the training data. For the smaller private and the Hillstrom data, we increased the number of trees to 2.000. 

To enable outcome adjustment, we also fit supervised learning models $\hat{\mu}(x)$, $\hat{\mu}_{0}(x)$ and $\hat{\mu}_{1}(x)$ (see section \ref{sec_adj_methods}) using the random forest implementation of the ranger package \citep{ranger}. 

\subsection{Assessment criteria}
We measure the impact of the outcome adjustment by comparing the $MSE_{W}$'s and the Qini curve's variance when calculated based on the adjusted outcome with the variance when calculated based on the original outcome. On the simulated data, we calculated the variance between the evaluation on the 10.000 test sets. On the real-world data sets, this is not possible because we just have a limited number of observations. Therefore, we calculated confidence intervals for the evaluation metrics on each of the real-world data sets (see \ref{app_CI} for details) and took the variance estimate of the confidence interval as an estimate of the metric's variance.

For the Qini curve, this variance can be calculated for each share $s$. On the real-world data sets, we calculated the variance at each decile ($s\in \{0.1,0.2,...\}$). On the simulation data, we only calculated the variance at the point $s=0.1$, due to computational reasons. For the $MSE_{W}$, we calculated differences
\begin{align*}
\Delta MSE_{W}(\hat{\tau}^{CF}_{x},\hat{\tau}^{triv}_{x})=MSE_{W}(\hat{\tau}^{CF}_{x})-MSE_{W}(\hat{\tau}^{triv}_{x}).
\end{align*} between the causal random forest $\hat{\tau}^{CF}_{x}$ and the trivial prediction $\hat{\tau}^{triv}_{x}=0$. Because outcome adjustment changes the expected value of $MSE_{W}(\hat{\tau}^{CF}_{x})$ but does not affect the expected value of $\Delta MSE_{W}$, comparing the variances of $\Delta MSE_{W}$ between adjusted and original outcome is more meaningful.      

In addition to the variance reduction, we calculated a second evaluation quality measure: For the $MSE_{W}$ on the simulated data, we calculated the "share of misleading evaluations". Thereby, we considered an evaluation misleading if $\Delta MSE_{W}(\hat{\tau}^{(1)}_{x},\hat{\tau}^{(2)}_{x})>0$, although the theoretical performance of $\hat{\tau}^{(1)}_{x}$ is better than the performance of $\hat{\tau}^{(2)}_{x}$. We calculated the share of misleading evaluations for comparisons of $\hat{\tau}^{CF}_{x}$ with the trivial predictions $\hat{\tau}^{triv}_{x}$, the perfect predictions $\hat{\tau}_{x}=\tau_{x}$, and the original causal random forest with some added noise $\hat{\tau}^{worse}_{x}\sim N(\hat{\tau}^{CF}_{x}, 0.01\sigma^{2}_{CF})$, where $\sigma_{CF}$ is the standard deviation in the predictions of $\hat{\tau}^{CF}_{x}$. In fact, misleading predictions are a consequence of the variance of an evaluation metric. Hence, measuring a reduction of the variance already guarantees that the probability of misleading evaluations decreases. But we think it is worth considering the share of misleading evaluations in addition to the percentage of variance reduction because it allows for a better understanding of the impact of variance reduction.                   

\subsection{Results}
\subsubsection{Variance reduction}
Tables \ref{tab_varred_sim} and \ref{tab_varred_real} show the percentage of variance reduction for the adjusted outcome-based metrics compared to the original outcome-based metrics. Notably, in each considered scenario (be it simulated data or real-world data) and for each metric, we observe a reduction in variance. This confirms our theoretical results.

For the $MSE_{W}$, unconditional mean adjustment leads to a sizeable variance reduction but, as expected, the variance reduction of conditional mean adjustment and doubly-robust adjustment is stronger in each scenario. For the simulated data, the variance reduction by conditional mean adjustment and doubly-robust adjustment surpasses even 90\% in some settings. For the real-world data, the reduction is smaller but still ranges between 12.4\% to 34.1\%. Likely, the gap between simulation settings and real-world data is due to the smaller impact of noise in the simulation. We can see that the variance reduction decreases with an increase in the noise standard deviation $\sigma$. Increasing $\sigma$ from 2 to higher values would probably yield a variance reduction comparable in magnitude to what we observe in the real-world data settings.

For the Qini curve, we generally notice a smaller variance reduction. This is again in line with our theory. For the Qini curve, outcome adjustment only reduces the variance of component \eqref{eq_var_comp_e} but, in contrast to the $MSE_{W}$, not of component \eqref{eq_var_comp_v}. The variance reduction of the Qini curve is always stronger than 10\% except for the Hillstrom data set, where the improvement is marginal. This shows that in real-world applications we would always expect to achieve a variance reduction but there is no guarantee that this reduction is sufficiently large to be practically relevant. The small improvement on the Hillstrom data set can be explained in the following way: On this data set, there is high treatment effect heterogeneity (variance in $\tau_{x}$), which is why it is possible to build useful uplift models on the data, but low hetorogeniety in the expected outcome (variance in $\mu_{x}$). As the variance in $\mu_{x}$ drives the reducible variance in the uplift metrics, there is little potential for improvement. 

In summary, conditional mean and doubly-robust adjustment reliably lead to a variance reduction of the $MSE_{W}$ and the Qini curve. The practical implications are easy to see when we translate the variance reduction into a reduction of the required test set sample size. For example, on the Criteo data set, applicants could save more than 10\% of the test set data and still benefit from the same confidence in their model evaluations if they apply outcome adjustment. As RCT data is costly to obtain, a substantial reduction of the required (test) sample size has practical relevance. 

The results of tables \ref{tab_varred_sim} and \ref{tab_varred_real} already provide clear evidence in favor of the use of outcome adjustment. However, we think that to better understand the rather abstract concept of variance reduction, some visualizations are helpful. Hence, we close this section by referring to Figures \ref{fig:mse_aw}, \ref{fig:rank_ai1} and \ref{fig:qini_bat}, which illustrate the results from table \ref{tab_varred_sim} and \ref{tab_varred_real}. The figures demonstrate the effect of outcome adjustment from a different perspective. For example, \ref{fig:qini_bat} shows the Qini curve calculated on the private data set by the original outcome and the adjusted outcomes. It demonstrates how variance reduction translates into smaller confidence intervals and additionally provides the variance reduction at each decile. Plots of the Qini curve for the other real-world data sets as well as plots of the distribution of $\Delta MSE_{W}$ are provided by \citet{bokelmab}.   
\begin{table*}[hbt]
\begin{center}
\caption{Variance reduction on simulated data}\label{tab_varred_sim}
\begin{threeparttable}

\begin{tabular*}{\textwidth}{p{0.08\textwidth}p{0.07\textwidth}|p{0.14\textwidth}p{0.14\textwidth}p{0.14\textwidth}p{0.13\textwidth}p{0.13\textwidth}}
    \toprule
    \textbf{Setting} & $\bm{\sigma}$ & $\bm{\Delta MSE_{W}(uc)}$ & $\bm{\Delta MSE_{W}(cond)}$ & $\bm{\Delta MSE_{W}(dr)}$  & \textbf{Qini\textsubscript{0.1}(cond)} & \textbf{Qini\textsubscript{0.1}(dr)} \\
    \hline\hline
    'aw' & 0.5  & 89.7\% & 97.8\% & 97.8\% & 10.1\% & 11.7\%  \\
    \hline
    'aw' & 1  & 83.0\% & 91.3\% & 91.4\% & 30.6\% & 31.3\%  \\
    \hline
    'aw' & 2  & 62.9\% & 69.8\% & 70.0\% & 14.3\% & 15.1\%  \\
    \hline
    'nw' & 0.5  & 60.3\% & 94.4\% & 93.8\% & 71.9\% & 69.5\%  \\
    \hline
    'nw' & 1  & 51.6\% & 80.8\% & 80.6\% & 47.5\% & 46.9\%  \\
    \hline
    'nw' & 2  & 28.4\% & 44.4\% & 45.0\% & 17.0\% & 17.8\%  \\
    \hline
     
\end{tabular*}

\begin{tablenotes}[para,flushleft]
   \small
   The table shows the percentage of variance reduction of the adjustment methods compared to the original outcome. The difference in $MSE_{W}$ was calculated between $\hat{\tau}_{x}^{CF}$ and $\hat{\tau}_{x}^{triv}$. For the Qini curve, the table shows the variance reduction at point $s=0.1$.    
   \end{tablenotes}

\end{threeparttable}
\end{center}
\end{table*}

\begin{table*}[hbt]
\begin{center}

\caption{Variance reduction on real-world data}\label{tab_varred_real}
\begin{threeparttable}

\begin{tabular*}{\textwidth}{p{0.13\textwidth}| p{0.15\textwidth}p{0.16\textwidth}p{0.15\textwidth}p{0.15\textwidth}p{0.15\textwidth}}
    \toprule
    \textbf{data}  & $\bm{\Delta MSE_{W}(uc)}$ & $\bm{\Delta MSE_{W}(cond)}$ & $\bm{\Delta MSE_{W}(dr)}$  & \textbf{Qini(cond)} & \textbf{Qini(dr)} \\
    \hline\hline
    private  & 16.4\% & 34.1\% & 33.3\% & 15.6-25.1\% & 15.1\%-23.8\%  \\
    \hline 
    Criteo  & 0.4\% & 13.8\% & 12.4\% & 12.7-13.7\% & 11.3-12.1\%  \\
    \hline
    Hillstrom  & 14.4\% & 16.7\% & 16.6\% & 1.79-3.30\% & 1.75-3.19\%  \\
    \hline
     
\end{tabular*}

\begin{tablenotes}[para,flushleft]
   \small
   The table shows the percentage of variance reduction of the adjustment methods compared to the original outcome. The difference in $MSE_{W}$ was calculated between $\hat{\tau}_{x}^{CF}$ and $\hat{\tau}_{x}^{triv}$. For the Qini curve, the table shows highest and lowest variance reduction over the 10 deciles, the Qini curve was claculated on.    
   \end{tablenotes}

\end{threeparttable}
\end{center}
\end{table*}

\subsubsection{Reduction of misleading evaluations}
Table \ref{tab_res_MSE}, shows the share of misleading evaluations by $MSE_{W}$ on the simulated data. In some scenarios, the results are quite remarkable: For "aw" with $\sigma=0.5$, the $MSE_{W}$ based on the original outcome mistakenly judges $\hat{\tau}^{CF}_{x}$ to perform better than the perfect prediction $\hat{\tau}_{x}=\tau_{x}$ in 39.3\% of the cases. The conditional mean or doubly-robust outcome adjustment reduces this share of misleading evaluations to 4.5\% and 4.4\%, respectively. On one hand, this demonstrates the high degree of randomness in model evaluations if the test set is not large enough. On the other hand, the result also displays the big impact that outcome adjustment can have on the quality of evaluations. Overall, the share of misleading evaluations is reduced by any of the adjustment methods in any setting. Thereby, as expected, conditional mean adjustment and doubly-robust adjustment perform better than unconditional mean adjustment.     

\begin{table*}[hbt]
\begin{center}
\caption{Share of misleading evaluations}\label{tab_res_MSE}
\begin{threeparttable}

\begin{tabular*}{\textwidth}{p{0.11\textwidth}p{0.11\textwidth}p{0.11\textwidth}|p{0.12\textwidth}p{0.12\textwidth}p{0.12\textwidth}p{0.12\textwidth}}
    \toprule
    \textbf{Setting} & $\bm{\hat{\tau}^{'}_{x}}$ & $\bm{\sigma}$ & \textbf{orig} & \textbf{cond} & \textbf{uncond} & \textbf{dr} \\
    \hline\hline
    "aw" & $\tau_{x}$ & $0.5$  & 39.3 & 4.5 & 23.6
 & \textbf{4.4} \\
    \hline
    "aw" & $\tau_{x}$ & $1$  & 32.9
 & 9.0 & 17.3 & \textbf{8.7} \\
    \hline
    "aw" & $\tau_{x}$ & $2$  & 25.1 & 9.4 & 11.2 & \textbf{9.2} \\
    \hline\hline
    "aw" & $\hat{\tau}^{triv}_{x}$ & $0.5$  & 6.1 & \textbf{0.0} & \textbf{0.0} & \textbf{0.0} \\
    \hline
    "aw" & $\hat{\tau}^{triv}_{x}$ & $1$  & 7.6 & \textbf{0.0} & \textbf{0.0} & \textbf{0.0} \\
    \hline
    "aw" & $\hat{\tau}^{triv}_{x}$ & $2$  & 9.9 & \textbf{0.7} & 1.3 & \textbf{0.7} \\
    \hline\hline
    "aw" & $\hat{\tau}^{worse}_{x}$ & $0.5$  & 48.6 & 39.0 & 44.7 & \textbf{38.8} \\
    \hline
    "aw" & $\hat{\tau}^{worse}_{x}$ & $1$  & 47.7 & 43.6 & 45.4
 & \textbf{43.4} \\
    \hline 
    "aw" & $\hat{\tau}^{worse}_{x}$ & $2$  & 48.5 & \textbf{46.4} & 46.6
 & \textbf{46.4} \\
    \hline\hline 
    "nw" & $\tau_{x}$ & $0.5$ 
 & 33.8 & \textbf{5.6} & 26.1
 & 6.6 \\
    \hline
    "nw" & $\tau_{x}$ & $1$  & 26.4
 & \textbf{8.0} & 19.6 & 8.2 \\
    \hline
    "nw" & $\tau_{x}$ & $2$  & 15.9
 & \textbf{10.7} & 13.0 & \textbf{10.7} \\
    \hline\hline
    "nw" & $\hat{\tau}^{triv}_{x}$ & $0.5$  & 16.8
 & \textbf{0.0} & 7.0 & \textbf{0.0} \\
    \hline
    "nw" & $\hat{\tau}^{triv}_{x}$ & $1$  & 28.5
 & 11.9 & 20.8 & \textbf{11.8} \\
    \hline
    "nw" & $\hat{\tau}^{triv}_{x}$ & $2$  & 56.9
 & \textbf{40.8} & 41.7 & \textbf{40.8} \\
    \hline\hline
    "nw" & $\hat{\tau}^{worse}_{x}$ & $0.5$  & 46.3
 & \textbf{38.8} & 44.7 & 38.9 \\
    \hline
    "nw" & $\hat{\tau}^{worse}_{x}$ & $1$ 
 & 46.7
 & 43.8 & 44.8 & \textbf{43.7}\\
 \hline
 "nw" & $\hat{\tau}^{worse}_{x}$ & $2$ & 45.6
 & 44.4 & 44.9 & \textbf{44.2}\\
    \hline
     
\end{tabular*}

\begin{tablenotes}[para,flushleft]
   \small
   The table shows the percentage of misleading evaluations, that is the share of simulation runs, where the MSE for $\hat{\tau}^{CF}_{x}$ was lower than the MSE for $\tau_{x}$, higher then for $\hat{\tau}^{triv}_{x}$ or higher than for $\hat{\tau}^{worse}_{x}$.     
   \end{tablenotes}

\end{threeparttable}
\end{center}
\end{table*}


\section{Conclusion}
In this paper, we showed that the statistical properties of the uplift evaluation metrics depend on the variance of the transformed outcome $W^{p}Y$. We also demonstrated that the variance of the transformed outcome can be reduced by fitting appropriate models $\hat{\Phi}(X)$ on the training set and adjusting the outcome on the test set to $Y-\hat{\Phi}(X)$. Importantly, our analysis confirmed that a corresponding adjustment leaves evaluation metrics unbiased and only reduces their variance. This reduced variance leads to a more reliable evaluation of uplift models according to the $MSE_{W}$ and also according to the Qini curve. For example, we found that an outcome adjustment produces far fewer misleading model comparisons by the $MSE_{W}$.

We described three possible outcome adjustment techniques, which we derived from the theoretical analysis. The unconditional mean adjustment leads to a reduction of variance for the $MSE_{W}$, but leaves the Qini curve unaffected. The conditional mean adjustment and the doubly-robust adjustment reduce the variance of the $MSE_{W}$ as well as the Qini curve. In all considered empirical settings, the  conditional mean adjustment and the doubly-robust adjustment reduced the variance of the metrics more than the unconditional outcome adjustment.  

For practical applications, we suggest to never use the $MSE_{W}$ with the original outcome. The results of our empirical analysis indicate that conclusions based on the $MSE_{W}$ with the original outcome could be severely misleading. Further, our theoretical analysis and empirical results clearly show that it is generally better to use adjusted outcomes. The easiest adjustment method is the unconditional mean adjustment because it does not require fitting a supervised learning model. But as we have seen, it is also inferior to the alternative adjustment approaches in form of the conditional mean adjustment or doubly-robust adjustment.  

Practical consequences of the variance reduction can be seen in two ways: If applicants evaluate uplift models on a given test set, evaluations based on the adjusted outcomes are more reliable than evaluations based on the original outcomes. Applicants would benefit from smaller confidence intervals due to our adjustments. Another benefit concerns the sample sizes needed to achieve a certain level of reliability (width of confidence interval). We have shown that variance reduction can translate into a reduction of the required sample size. Our empirical analysis shows that a reduction of the required test sample size of 10\% and more are realistic in practice. 

In this paper, we focused on the $MSE_{W}$ and the Qini curve as empirical uplift evaluation metrics. In the Appendix, we also summarised the other metrics, which appear in the uplift literature. We theoretically show that the principle of variance reduction by outcome adjustment also applies there. In fact, we are not aware of a form of uplift model evaluation that would not benefit from  outcome adjustment. However, the magnitude of improvement depends on the specific data set and may be small. 

In summary, the suggested outcome adjustment methods offer clear practical benefits in terms of variance reduction. They also do not bring notable disadvantages: they are easy to apply, can not cause bias, and will almost surely reduce variance unless the required supervised learning models predict worse than a naive model. For these reasons, we think it is worth changing current state-of-art uplift evaluation procedures towards using adjusted outcomes. 

\appendix

\section{Derivation of the metric distributions}
\subsection{Derivation of the transformed mean squared error distribution}\label{app_distr_TOMSE}

Given the outcome representation in equation \eqref{eq_dec_Y}, we can write the transformed outcome as
\begin{align*}
W^{p}Y=\tau_{x}+\zeta,
\end{align*} with $\zeta:=W^{p}(\mu_{x}+(1-p)\tau_{x})+W^{p}\varepsilon$. The difference in squared deviation from $W^{p}Y$ between two estimators $\hat{\tau}^{(1)}_{x}, \hat{\tau}^{(2)}_{x}$ then becomes
\begin{align*}
(W^{p}Y-\hat{\tau}_{x}^{(1)})^{2}-(W^{p}Y-\hat{\tau}_{x}^{(2)})^{2}&=(\tau_{x}+\zeta-\hat{\tau}^{(1)}_{x})^{2}-(\tau_{x}+\zeta-\hat{\tau}^{(2)}_{x})^{2}\\
&=(\tau_{x}-\hat{\tau}^{(1)}_{x})^{2}-(\tau_{x}-\hat{\tau}^{(2)}_{x})^{2}+2\zeta(\hat{\tau}^{(2)}_{x}-\hat{\tau}^{(1)}_{x}).
\end{align*}
Using the fact that $E[\zeta|X]=0$ and $\hat{\tau}_{x}^{(1)},\hat{\tau}_{x}^{(2)},\tau_{x}$ are functions in $X$, we can see that
\begin{align*}
E[(W^{p}Y-\hat{\tau}_{x}^{(1)})^{2}-(W^{p}Y-\hat{\tau}_{x}^{(2)})^{2}|X]&=(\tau_{x}-\hat{\tau}^{(1)}_{x})^{2}-(\tau_{x}-\hat{\tau}^{(2)}_{x})^{2} \text{ and}\\
Var[(W^{p}Y-\hat{\tau}_{x}^{(1)})^{2}-(W^{p}Y-\hat{\tau}_{x}^{(2)})^{2}|X]&=4(\hat{\tau}^{(2)}_{x}-\hat{\tau}^{(1)}_{x})^{2}Var[\zeta|X].
\end{align*}

In this way, we can derive equations for the expected value and variance of the $MSE_{W}$. We denote the difference in $MSE_{W}$ by $\Delta MSE_{W}(\hat{\tau}_{x}^{(1)},\hat{\tau}_{x}^{(2)}):=MSE_{W}(\hat{\tau}_{x}^{(1)})-MSE_{W}(\hat{\tau}_{x}^{(2)})$. Note that the $MSE_{W}$ is an average of $N$ squared deviations of the estimators from $W^{p}Y$. The above equations then become 
\begin{align}
E[\Delta MSE_{W}(\hat{\tau}_{x}^{(1)},\hat{\tau}_{x}^{(2)})|X]&=(\tau_{x}-\hat{\tau}^{(1)}_{x})^{2}-(\tau_{x}-\hat{\tau}^{(2)}_{x})^{2} \text{ and}\label{eq_exp_delta_mse}\\
Var[\Delta MSE_{W}(\hat{\tau}_{x}^{(1)},\hat{\tau}_{x}^{(2)})|X]&=\frac{4(\hat{\tau}^{(2)}_{x}-\hat{\tau}^{(1)}_{x})^{2}}{N}Var[\zeta|X].
\end{align}
Equation \eqref{eq_exp_delta_mse} shows that $MSE_{W}$ is unbiased. Now, we examine the variance. Given the law of total variance, we obtain
\begin{align*}
Var[\Delta MSE_{W}(\hat{\tau}_{x}^{(1)},\hat{\tau}_{x}^{(2)})]=E[Var[\Delta MSE_{W}(\hat{\tau}_{x}^{(1)},\hat{\tau}_{x}^{(2)})|X]]+Var[E[\Delta MSE_{W}(\hat{\tau}_{x}^{(1)},\hat{\tau}_{x}^{(2)})|X]].
\end{align*} From equation \eqref{eq_exp_delta_mse}, we see that the second term does not depend on the outcome $Y$ and is hence not reducible. Accordingly, the reducible part of the variance is
\begin{align*}
R:=\frac{4}{N}E[(\hat{\tau}^{(2)}_{x}-\hat{\tau}^{(1)}_{x})^{2}Var[W^{p}Y|X]].
\end{align*}
Some transformations yield
\begin{align*}
\frac{N}{4}R=Cov((\hat{\tau}^{(2)}_{x}-\hat{\tau}^{(1)}_{x})^{2},Var[W^{p}Y|X])+E[(\hat{\tau}^{(2)}_{x}-\hat{\tau}^{(1)}_{x})^{2}]\cdot E[Var[W^{p}Y|X]].
\end{align*}
Minimizing this irrespective of the choice of $\hat{\tau}^{(1)}_{x}$ and $\hat{\tau}^{(2)}_{x}$, which is a plausible approach as the metric should be optimal for the evaluation of all estimators, corresponds to minimizing $E[Var[W^{p}Y|X]]$. To do this, we again apply the law of total variance
\begin{align*}
E[Var[W^{p}Y|X]]&=Var[W^{p}Y]-Var[E[W^{p}Y|X]]\\
&=Var[W^{p}Y]-Var[\tau_{x}].
\end{align*} So, in summary, a reduction of $Var[\Delta MSE^{\hat{\Phi}}_{W}(\hat{\tau}_{x}^{(1)},\hat{\tau}_{x}^{(2)})]$ corresponds to a reduction of $Var[W^{p}Y]$. 

\begin{figure}[!htb]
\centering
  \includegraphics[width=0.6\textwidth]{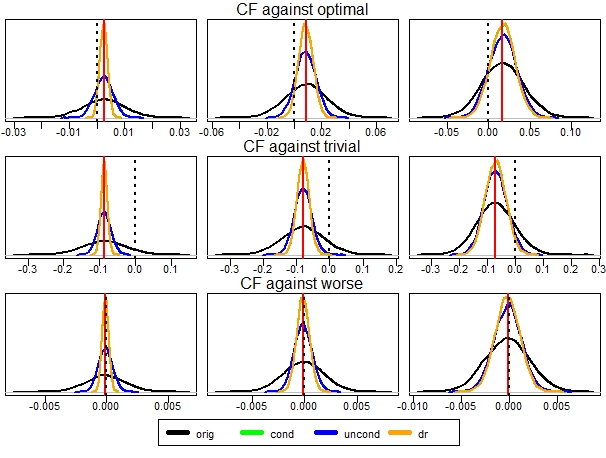}
  \small
  \captionof{figure}{Results for the $MSE_{W}$ in the setting "aw". The plots show the density of differences in MSE between causal random forest $\hat{\tau}^{CF}_{x}$ and optimal predictions $\tau_{x}$ (first row), trivial predictions $\hat{\tau}^{triv}_{x}$ (second row) and causal random forest predictions with added noise $\hat{\tau}^{worse}_{x}$ (third row). The considered setting is "aw". For the pictures in the left column, the noise standard deviation was set to $\sigma = 0.5$, for the middle column to $\sigma = 1$, and for the right column to $\sigma = 2$.}
  \label{fig:mse_aw}
\end{figure}

\subsection{Derivation of the Qini metric distribution}\label{ap_qini_distr}
The estimate $\hat{ATE}_{s}$ of the average treatment effect for the share $s$ of highest ranked individuals is
\begin{align*}
\hat{ATE}_{s}(\hat{\tau}_{x})=\frac{1}{N_{W}}\sum Y_{i}|_{W=1}-\frac{1}{N_{\bar{W}}}\sum Y_{i}|_{W=0},
\end{align*} where the outcomes $Y_{i}$ belong to individuals ranked in the highest share $s$ and $N_{W}$, $N_{\bar{W}}$ denote the amount of treated and untreated individuals within this share, respectively. This expression can be transformed into 
\begin{align*}
\hat{ATE}_{s}(\hat{\tau}_{x})&=\frac{1}{N}\sum W_{i}^{\tilde{p}}Y_{i},
\end{align*} where $N=N_{W}+N_{\bar{W}}$ and $\tilde{p}=\frac{N_{W}}{N}$.

This estimate is unbiased:
\begin{align*}
E[\hat{ATE}_{s}(\hat{\tau}_{x})]&=\frac{1}{N}\sum E[W_{i}^{\tilde{p}}Y_{i}]\\
&=E[\tau_{x}],
\end{align*} where the CATE $\tau_{x}$ belongs to individuals ranked in the highest share $s$. Next, we analyze the variance. By the law of total variance, we can derive
\begin{align*}
Var[\frac{1}{N}\sum W_{i}^{\tilde{p}}Y_{i}]&=\frac{1}{N}Var[W_{i}^{\tilde{p}}Y_{i}]\\
&=\frac{1}{N}(E[Var[W_{i}^{\tilde{p}}Y_{i}|(W_{i})_{i=1,...,N}]]+Var[E[W_{i}^{\tilde{p}}Y_{i}|(W_{i})_{i=1,...,N}]])\\
&=\frac{1}{N}E[Var[W^{\tilde{p}}Y|W]].
\end{align*} In this derivation, $Var[E[W_{i}^{\tilde{p}}Y_{i}|(W_{i})_{i=1,...,N}]]=0$ follows from the definition of $\tilde{p}=\frac{\sum W_{i}}{N}$, which makes $E[W_{i}^{\tilde{p}}Y_{i}|(W_{i})_{i=1,...,N}]=E[\tau_{x}]$ constant irrespective of $(W_{i})_{i=1,...,N}$. Thus, the variance of $\hat{ATE}_{s}(\hat{\tau}_{x})$ is 
\begin{align*}
Var[\hat{ATE}_{s}(\hat{\tau}_{x})]=\frac{1}{N}E[Var[W^{\tilde{p}}Y|W]],
\end{align*} where we keep in mind that the outcome $Y$ on the right side of the equation belongs to individuals in the share $s$ with the highest rank. 

\begin{figure}[!htb]
\centering
  \includegraphics[width=0.6\textwidth]{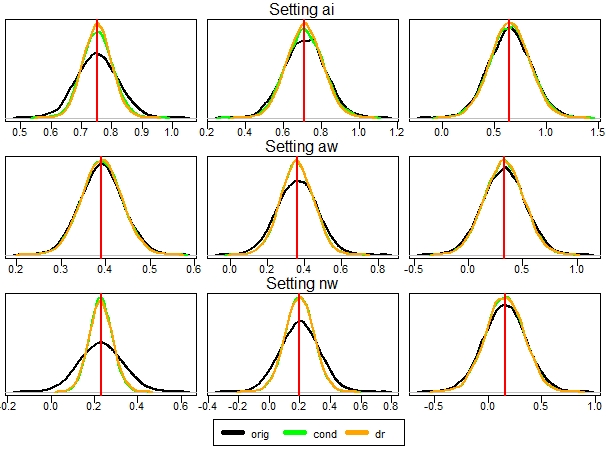}
  \small
  \captionof{figure}{Density of estimators for $ATE_{0.1}(\hat{\tau}_{x})$ by causal random forest. For the pictures in the left column, the noise standard deviation was set to $\sigma = 0.5$, for the middle column to $\sigma = 1$ and for the right column to $\sigma = 2$.}
  \label{fig:rank_ai1}
\end{figure}

\subsection{Calculation of confidence intervals}\label{app_CI}
The distribution of an evaluation metric determines the confidence intervals. For unbiased error metrics like the $MSE_{W}$ and the Qini curve, the variance translates into the width of the confidence interval. Next, we derive formulas for the confidence intervals of the $MSE_{W}$ and the Qini curve. 

For the Qini curve, we can construct a confidence interval at each point $s$ by analyzing the distribution of $\hat{ATE}_{s}(\hat{\tau}_{x})$ around the theoretical value $ATE_{s}$. Given equation (\ref{eq_qini}) for the original outcome, we can assume asymptotic normality with
\begin{align*}
\hat{ATE}_{s}&\sim N(ATE_{s},\sigma_{s}^{2}) \text{ and}\\
\sigma_{s}^{2}&:=\frac{Var[Y|W=1]}{N_{W}}+\frac{Var[Y|W=0]}{N_{\bar{W}}},
\end{align*} where the outcome $Y$ belongs to the share $s$ of highest ranked individuals. As the values of the Qini curve are calculated as $u_{s}=\hat{ATE}_{s}\cdot N_{W}(s)$, the 95\% confidence interval at each point can be constructed by 
\begin{align*}
CI_{s}=\left[\left(\hat{ATE}_{s}-1.96\sqrt{\frac{\hat{\sigma}_{s,W}^{2}+\hat{\sigma}_{s,\bar{W}}^{2}}{N(s)}}\right)\cdot N_{W}(s),\left(\hat{ATE}_{s}+1.96\sqrt{\frac{\hat{\sigma}_{s,W}^{2}+\hat{\sigma}_{s,\bar{W}}^{2}}{N(s)}}\right)\cdot N_{W}(s)\right],
\end{align*} where $\hat{\sigma}_{s,W}, \hat{\sigma}_{s,\bar{W}}$ are the empirical estimates for $Var[Y|W=1]$ and $Var[Y|W=0]$ and $N(s)$ is the total number of individuals within the share $s$ with highest rank. In the same way, we can calculate confidence intervals for the adjusted outcomes.

For the $MSE_{W}$, it is of little use to compare confidence intervals between the original and the adjusted outcomes because outcome adjustment changes the expected value of $MSE_{W}$. Confidence intervals for the different outcome versions would thus be built around different parameters. Instead, it is plausible to calculate confidence intervals for a difference in $MSE_{W}$ because for differences, outcome adjustment does not change the expected value. The difference in $MSE_{W}$ can be written as
\begin{align*}
\Delta MSE_{W}(\hat{\tau}_{x}^{(1)}, \hat{\tau}_{x}^{(2)})=\frac{1}{N}\sum \left[(W^{p}Y_{i}-\hat{\tau}_{x_{i}}^{(1)})^{2} - (W^{p}Y_{i}-\hat{\tau}_{x_{i}}^{(2)})^{2}\right].
\end{align*} As a mean of independent summands, $\Delta MSE_{W}(\hat{\tau}_{x}^{(1)}, \hat{\tau}_{x}^{(2)})$ is asymptotically normally distributed. Hence, a confidence interval can be calculated by
\begin{align*}
CI=\left[\Delta MSE_{W}(\hat{\tau}_{x}^{(1)}, \hat{\tau}_{x}^{(2)})-1.96\frac{\hat{\sigma}}{\sqrt{N}},\Delta MSE_{W}(\hat{\tau}_{x}^{(1)}, \hat{\tau}_{x}^{(2)})+1.96\frac{\hat{\sigma}}{\sqrt{N}}\right],
\end{align*} where $\hat{\sigma}$ is the empirical standard deviation between the summands $\left[(W^{p}Y_{i}-\hat{\tau}_{x_{i}}^{(1)})^{2} - (W^{p}Y_{i}-\hat{\tau}_{x_{i}}^{(1)})^{2}\right]$.

\begin{figure}[!htb]
\centering
  \includegraphics[width=0.6\textwidth]{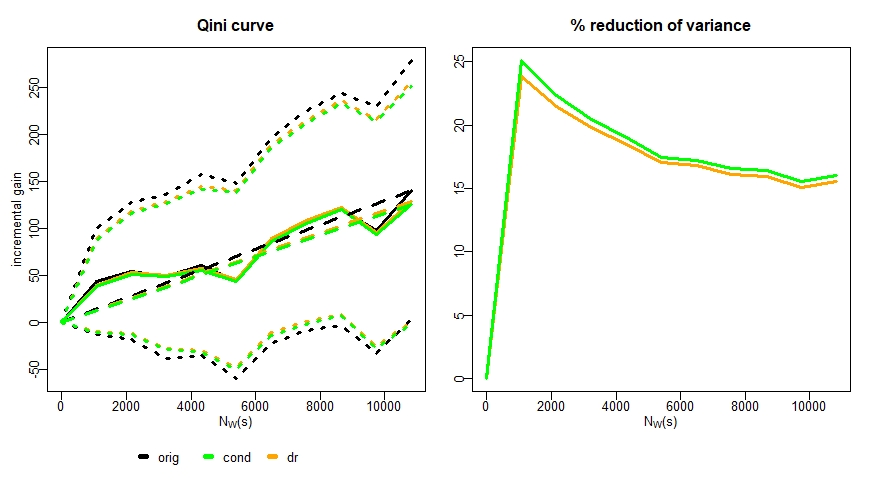}
  \small
  \captionof{figure}{Left panel: Qini curve calculated by the original outcome, the conditional mean adjusted outcome, and the doubly-robust adjusted outcome. The dashed lines represent the confidence intervals. Right panel: Percentage reduction of variance by the adjusted outcomes methods compared to the original outcome.}
  \label{fig:qini_bat}
\end{figure}

\section{Variance components of the transformed outcome}

\subsection{Calculation of the variance components}\label{app_calc_var_comp}  

In section \ref{sec_met_var_red}, \ref{app_distr_TOMSE}, and \ref{ap_qini_distr}, we have examined the variance of the uplift evaluation metrics and have found that the variance of the $MSE_{W}$ depends on both of the components $E[Var[W^{p}Y|W]]$ and $Var[E[W^{p}Y|W]]$. We have further found that the variance of the Qini curve only depends on the first of these components, with the slight modification that the treatment probability $p$ in the RCT needs to be replaced by the fraction of treated $\tilde{p}$ within the share $s$ of highest ranked individuals.

Here, we further decompose the variance components to reveal their relationship to the nuisance function $\Phi(X)$. It holds
\begin{align*}
E[Var[W^{p}Y|W]]&=E[Var[\tau_{x}+W^{p}\Phi(X)+W^{p}\varepsilon|W]]\\
&=E[Var[\tau_{x}]+W^{p}Cov(\tau_{x},\Phi(X))+(W^{p})^{2}Var[\Phi(X)]+(W^{p})^{2}Var[\varepsilon]]\\
&=Var[\tau_{x}]+\frac{Var[\Phi(X)]+Var[\varepsilon]}{p(1-p)} \text{ and}\\
\\
Var[E[W^{p}Y|W]]&=Var[E[\tau_{x}+W^{p}\Phi(X)+W^{p}\varepsilon|W]]\\
&=Var[E[\tau_{x}]+W^{p}E[\Phi(X)]]\\
&=\frac{E[\Phi(X)]^{2}}{p(1-p)}.
\end{align*}

To reduce the variance of the transformed outcome, we need to reduce $Var[\Phi(X)]$ and $E[\Phi(X)]^{2}$. Note that the latter term only affects the transformed outcome for the MSE, while the former affects also the transformed outcome for the rank metrics.

\subsection{Reduction of the variance components}\label{app_red_vc}
In this section, we show that a supervised learning model $\hat{\Phi}(X)$ for the target $\Phi(X)$ will generally fulfill conditions \eqref{eq_cond1} and \eqref{eq_cond2} for variance reduction. We start with condition \eqref{eq_cond1}. The mean squared error, which is the plausible optimization criterion for $\hat{\Phi}(X)$, can be written as
\begin{align*}
E[(\Phi(X)-\hat{\Phi}(X))^{2}]=Var[\Phi(X)-\hat{\Phi}(X)]+(E[\Phi(X)]-E[\hat{\Phi}(X)])^{2}.
\end{align*} Assume condition \eqref{eq_cond1} is not fulfilled. Then the difference in expected values, in the second component, would be larger than $|E[\Phi(X)]|$ in expected value. In case the difference in expected values is positive and we shift our model by a constant $c\in (0,2|E[\Phi(X)]|)$, the second component would be reduced. In case the difference in expected values is negative and we choose $c\in (-2|E[\Phi(X)]|,0)$, the second component would again be reduced. Such a constant shift of the model $\hat{\Phi}(X)+c$ would not affect the first component on the right-hand side. Accordingly, such a shift would reduce the MSE. For a properly fitted model, it is impossible to miss such a simple shift during the training unless $E[\Phi(X)]$ is close to zero. So, if $E[\Phi(X)]$ is not close to zero, condition \eqref{eq_cond1} will be fulfilled.      
We next examine condition \eqref{eq_cond2}. If $\hat{\Phi}(X)$ is an estimator for $\Phi(X)$ that is more precise than the trivial $E[\Phi(X)]$ in terms of the MSE loss, then
\begin{align*}
Var[\Phi(X)-\hat{\Phi}(X)]&=E[(\Phi(X)-\hat{\Phi}(X))^{2}]-E[(\Phi(X)-\hat{\Phi}(X))]^{2}\\
&\leq E[(\Phi(X)-\hat{\Phi}(X))^{2}]\\
&< E[(\Phi(X)-E[\Phi(X)])^{2}]\\
&=Var[\Phi(X)].
\end{align*} So, any supervised learning model $\hat{\Phi}(X)$ for the parameter $\Phi(X)$ that predicts better than a naive model will fulfill condition \eqref{eq_cond2} for variance reduction. 

\section{Other evaluation metrics}\label{ap_other_metrics}
In our analysis of error metrics, we have focused on the Qini curve and the $MSE_{W}$, which are commonly used in the uplift literature. Here, we summarise other evaluation metrics found in the literature and show that our results also apply to these metrics. In general, there are three classes of empirical evaluation metrics: rank metrics, treatment decision metrics, which measure the impact of allocating treatment according to the uplift model, and accuracy metrics. The Qini curve belongs to the first class and the $MSE_{W}$ to the third. In the following, we will examine other members of these three metric classes.

\subsection{Rank metrics: Qini curve, uplift curve, and related measures}
In section \ref{sec_rank_metr}, we analyzed the Qini curve by \citet{radcliffe2007using}, as one of the most popular empirical ranking evaluation methods. This curve is given by
\begin{align}
u_{s}=\sum Y_{i}|_{W=1}-\frac{N_{W}(s)}{N_{\bar{W}}(s)}\sum Y_{i}|_{W=0}\label{qini1} \tag{Q1},
\end{align} where $N_{W}(s)$ and $N_{\bar{W}}(s)$ denote the number of treated respectively untreated in the share $s$ of highest ranked individuals. Note that here we use $N_{W}(s)$ in the notation and not $N_{W}$ to highlight that we only count the number of treated within the share $s$ and not in the whole data set. This is important to represent differences to other variants of the Qini curve. \citet{devriendt2020learning} provide an extensive review of alternative approaches to calculate the Qini curve. In the following, we briefly examine these variants and show why our statistical analysis for the original Qini curve also applies.

The first variation of the Qini curve is, for example, used by \citet{Diemert2018}. They calculate the Qini curve as
\begin{align*}
u_{s}=\sum Y_{i}|_{W=1}-\frac{N_{W}}{N_{\bar{W}}}\sum Y_{i}|_{W=0}.\label{qini2} \tag{Q2}
\end{align*} Another version of the Qini curve is provided, for example, by \citet{guelman2015optimal}, where it is defined by
\begin{align*}
u_{s}=\frac{1}{N_{W}}\sum Y_{i}|_{W=1}-\frac{1}{N_{\bar{W}}}\sum Y_{i}|_{W=0}.\label{qini3} \tag{Q3}
\end{align*} These alternative Qini curves fit our analysis of the statistical properties of the original Qini curve. For \eqref{qini1}, it holds $u_{s}=\hat{ATE}_{s}\cdot N_{W}(s)$. For \eqref{qini2}, it holds due to the random treatment allocation $N_{W}(s)\approx s\cdot N_{W}$ and $N_{\bar{W}}(s)\approx s\cdot N_{\bar{W}}$. This leads to $u_{s}\approx \hat{ATE}_{s} N_{W}(s)$. For \eqref{qini3}, it holds for similar reasons $u_{s}\approx \hat{ATE}_{s}\cdot s$. So, all the Qini curve versions correspond to $\hat{ATE}_{s}$ multiplied by a factor that is unrelated to the distribution of the outcome. This is exactly the same situation as for the original Qini curve. In the analysis of its statistical properties, we only needed to examine the properties of $\hat{ATE}_{s}$. And as we can see, these also determine the statistical properties of the other Qini curve versions.   

One "alternative" to the Qini curve is the so-called "uplift curve". In some studies like \citet{gutierrez2017causal}, the uplift curve is given by 
\begin{align}
\bar{u}_{s}=(\frac{1}{N_{W}(s)}\sum Y_{i}|_{W=1}-\frac{1}{N_{\bar{W}}(s)}\sum Y_{i}|_{W=0})\cdot (N_{W}(s)+N_{\bar{W}}(s))\label{uc1} \tag{UC1}.
\end{align} In other papers, like \citet{jaskowski2012uplift}, it is given by 
\begin{align*}
\bar{u}_{s}=\frac{1}{N_{W}}\sum Y_{i}|_{W=1}-\frac{1}{N_{\bar{W}}}\sum Y_{i}|_{W=0}\label{uc2} \tag{UC2}.
\end{align*} Another alternative is provided by \citet{kuusisto2014support}, who calculate the uplift curve by
\begin{align*}
\bar{u}_{s}=\sum Y_{i}|_{W=1}-\sum Y_{i}|_{W=0}\label{uc3} \tag{UC3}.
\end{align*} Again, these forms of the uplift curve fit our analysis of the statistical properties of the original Qini curve. For \eqref{uc1}, we can see that $\bar{u}_{s}=\hat{ATE}_{s}\cdot (N_{W}(s)+N_{\bar{W}}(s))$. For \eqref{uc2}, we can see that due to random treatment allocation $N_{W}(s)\approx s\cdot N_{W}$ and $N_{\bar{W}}(s)\approx s\cdot N_{\bar{W}}$. This then leads to $\bar{u}_{s}\approx \hat{ATE}_{s}\cdot s$. As to \eqref{uc3}, this metric only works of the treatment property if $p=0.5$. Then, $N_{W}(s)\approx N_{\bar{W}}(s)$ and, therefore, $\bar{u}_{s}\approx\hat{ATE}_{s}\cdot \frac{N_{W}(s)+N_{\bar{W}}(s)}{2}$. So, with the same argumentation as for the alternative Qini curves, we can see that the statistical analysis of the original Qini curve applies to different forms of uplift curves.

Another related way to empirically evaluate the ranking of a model is to visualize the "uplift per decile", as done by \citet{kane2014mining}. This works by estimating the average treatment effects $\hat{ATE}_{[0,0.1]}, \hat{ATE}_{[0,0.2]},...$ per decile as the difference in mean outcomes between the treated and untreated. Statistically, this is the same as what happens in equation \eqref{eq_ates}, just that the outcomes belong to individuals whose predictions are in a certain decile instead of the share $s$ of individuals with the highest rank. Therefore, we can also transfer the statistical analysis to this evaluation metric. 

In addition to the various versions of Qini and uplift curves, some evaluation measures are derived from these curves. An example is the \textit{area under the uplift curve} (AUUC), which \citet{devriendt2020learning} calculate as
\begin{align*}
AUUC\approx \sum_{k=1}^{100} \bar{u}_{k/100}.
\end{align*} Of course, the statistical properties of this measure depend on the statistical properties of the uplift curve $\bar{u}_{k/100}$. Since the statistical properties of the uplift curve can be analyzed in the same way as the statistical properties of the Qini curve by \citet{radcliffe2007using}, the results of our statistical analysis directly transfer to the area under the uplift curve.

Finally, \citet{radcliffe2007using} defines two \textit{Qini} values, which he derives from the Qini curve. The Qini values measure two versions of a ratio between the Qini curve and a theoretical optimum. Again, our statistical analysis holds for these measures because it concerns the statistical properties of the Qini curve.

In summary, we find out statistical analysis holds for a range of ranking evaluation measures used in the uplift modeling literature and different versions of these measures. This also implies that all of these measures and versions suffer from variance and will benefit from the proposed adjustments for variance reduction.

\subsection{Treatment decision metrics}
Instead of measuring how well a model ranks individuals according to $\tau_{x}$, it is also possible to measure what happens if we derive concrete treatment allocation decisions $d(X)$ from an uplift model. There are two theoretical parameters that would measure model performance: The \textit{gain} $E[\tau_{x}|d(X)=1]$ and the \textit{expected outcome} $E[Y|W=d(X)]$ \textit{if treatment allocation works according to the model decisions}.

The gain could be estimated by 
\begin{align*}
\hat{ATE}_{d=1}=\frac{1}{N_{W}}\sum Y_{i}|_{W=1}-\frac{1}{N_{\bar{W}}}\sum Y_{i}|_{W=0},
\end{align*} where the outcomes belong to individuals for which the model recommends treatment \citep{schuler2018comparison}. This is again an average treatment effect estimator by differences in the means between the treated and the untreated. So, just like in the analysis of the Qini curve  \citet{radcliffe2007using}, our results concerning variance reduction apply. 

The expected outcome if treatment is assigned according to uplift model recommendation can be estimated by
\begin{align*}
\hat{v}(d)=\frac{1}{N}(\sum\frac{W_{i}\cdot d(X_{i})}{\tilde{p}}Y_{i}+\sum\frac{(1-W_{i})(1-d(X_{i}))}{1-\tilde{p}}Y_{i}),
\end{align*}
where $\tilde{p}=\frac{\sum W_{i}}{N}$. This metric appears in the literature under the name "decision value" \citep{schuler2018comparison, kapelner2014inference, zhao2017uplift}. \citet{hitsch2018heterogeneous} use the same metric under the name "targeting profit", the only difference being that they added a cost $c$ for each decision to treat, which is irrelevant in terms of the statistical properties. $\hat{v}(d)$ is more challenging to discuss than any of the other empirical evaluation metrics described so far. This is because outcome adjustment would bias the metric (because of $E[Y-\hat{\Phi}(X)|W=d(X)]\neq E[Y|W=d(X)]$). But if we use this metric to evaluate which of two decision models $d_{1}(x), d_{2}(x)$ yields better decisions, the bias due to outcome adjustment disappears because 
\begin{align*}
E[Y-\hat{\Phi}(X)|W=d_{1}(X)]-E[Y-\hat{\Phi}(X)|W=d_{2}(X)]=&E[Y|W=d_{1}(X)]-E[\hat{\Phi}(X)|W=d_{1}(X)]\\
&-E[Y|W=d_{2}(X)]+E[\hat{\Phi}(X)|W=d_{2}(X)]\\
=&E[Y|W=d_{1}(X)]-E[Y|W=d_{2}(X)]
\end{align*} To derive that the variance can be reduced, we write the difference between the empirical estimates as
\begin{align*}
\hat{v}(d_{1})-\hat{v}(d_{2})=\frac{1}{N}(\sum_{\substack{d_{1}=1 \\ d_{2}=0}}W^{p}_{i}Y_{i}-\sum_{\substack{d_{1}=0 \\ d_{2}=1}}W^{p}_{i}Y_{i}).
\end{align*} We obtain sums of transformed outcomes $W^{p}_{i}Y_{i}$ in this expression. In section \ref{sec_met_var_red}, we have shown that outcome adjustment leads to a reduction of $Var[W^{p}_{i}Y_{i}]$. This would then also apply for the difference $\hat{v}(d_{1})-\hat{v}(d_{2})$.   

\subsection{Accuracy metrics}\label{app_alt_acc}
Here, we discuss alternative accuracy metrics beyond the $MSE_{W}$. The first two methods are described and evaluated by \citet{saito2020counterfactual} in the context of observational data. Here, we discuss them in the context of RCT data. The first method is the \textit{plug-in evaluation}. It measures the uplift model performance by
\begin{align*}
MSE_{pi}(\hat{\tau}_{x})=\frac{1}{N}\sum (\hat{\mu}_{1}(x_{i})-\hat{\mu}_{0}(x_{i})-\hat{\tau}_{x_{i}})^{2},
\end{align*} where $\hat{\mu}_{0}(x)$ and $\hat{\mu}_{1}(x)$ are estimators of the conditional expected value of the treated and untreated respectively. In an empirical analysis, \citet{saito2020counterfactual} found the $MSE_{pi}$ slightly inferior to the $MSE_{W}$ with doubly-robust adjustment. In our opinion, however, there is a more serious problem with the $MSE_{pi}$. It is easy to verify that $MSE_{pi}$ is biased. If we choose $\hat{\tau}_{x}:=\hat{\mu}_{1}(x)-\hat{\mu}_{0}(x)$ as a CATE estimator to evaluate, $MSE_{pi}(\hat{\tau}_{x})$ would be zero. 

The second metric described and evaluated by \citet{saito2020counterfactual} is called \textit{$\tau$-risk}. It is based on the loss function of the R-learner from \citet{nie2021quasi}. For RCT data, where the treatment probability $p$ is constant, it would be defined as
\begin{align*}
MSE_{\tau}(\hat{\tau}_{x})=\frac{1}{N}\sum (Y_{i}-\hat{\mu}(x_{i})-(W_{i}-p)\hat{\tau}_{x_{i}})^{2},
\end{align*} where $\hat{\mu}(x)$ is an estimator of the conditional expected value of treated and untreated combined. In the empirical analysis by \citet{saito2020counterfactual}, this metric does not perform well. Furthermore, just like the $MSE_{pi}$, this metric is also biased. We show this by using representation \eqref{eq_dec_Y} of the outcome in the definition of the metric:
\begin{align*}
MSE_{\tau}(\hat{\tau}_{x})&=\frac{1}{N}\sum (\mu_{x_{i}}+W_{i}\tau_{x_{i}}+\varepsilon-\hat{\mu}(x_{i})-(W_{i}-p)\hat{\tau}_{x_{i}})^{2}\\
&=\frac{1}{N}\sum (\mu_{x_{i}}-\hat{\mu}(x_{i})+p\cdot \hat{\tau}_{x_{i}}+W_{i}(\tau_{x_{i}}-\hat{\tau}_{x_{i}})+\varepsilon_{i})^{2}\\
&=\frac{1}{N}\sum (A_{x_{i}}+W_{i}(\tau_{x_{i}}-\hat{\tau}_{x_{i}})+\varepsilon_{i})^{2} \text{, with}\\
A_{x_{i}}&:=\mu_{x_{i}}-\hat{\mu}(x_{i})+p\cdot \hat{\tau}_{x_{i}}.
\end{align*} The expected value is given by $E[MSE_{\tau}(\hat{\tau}_{x})]=E[A^{2}_{x_{i}}]+pE[(\tau_{x_{i}}-\hat{\tau}_{x_{i}})^{2}]+E[\varepsilon^{2}_{i}]$. So, for an estimator with $\hat{\tau}_{x}:=\frac{\hat{\mu}(x)-\mu_{x}}{p}$, the term $E[A^{2}_{x_{i}}]$ is zero. Accordingly, this estimator would be preferred by the $MSE_{\tau}$, compared to an estimator with the same theoretical performance $E[(\tau_{x_{i}}-\hat{\tau}_{x_{i}})^{2}]$, but with $A_{x_{i}}\neq 0$.    

Another metric is the \textit{$\mu$-loss} described by \citet{schuler2018comparison}. It is only applicable to uplift models that  yield, for each individual, two outcome predictions: $\hat{\mu}_{1}(x_{i})$ if the individual is treated and $\hat{\mu}_{0}(x_{i})$ if the individual is not treated. The $\mu$-loss is then calculated by
\begin{align*}
MSE_{\mu}=\frac{1}{N}\sum(Y_{i}-W\hat{\mu}_{1}(x_{i})-(1-W)\hat{\mu}_{0}(x_{i}))^{2}.
\end{align*} 

In summary, accuracy evaluation metrics other than the $MSE_{W}$ are either biased or not generally applicable for CATE model evaluation. In our opinion, this renders the corresponding metrics unsuitable for the evaluation of uplift models on RCT data. Notably, bias disqualifies the metrics mainly for application to RCT. For observational data, any metric could be biased due to the unknown treatment probability $p(x)$. In contrast, on RCT data it is easy to build useful and unbiased evaluation metrics. Hence, any biased metric can be considered unsuitable. Accordingly, we would recommend to use the $MSE_{W}$ (with our suggested outcome adjustment methods) to evaluate the accuracy of uplift model predictions on RCT data.  

\bibliographystyle{unsrtnat}
\bibliography{template}







\end{document}